# The fundamental problem of risk prediction for individuals: health AI, uncertainty, and personalized medicine

**Authors**: Lasai BARREÑADA[1,2], Ewout STEYERBERG[3,4], Dirk TIMMERMAN[1,5], Doranne THOMASSEN[3], Laure WYNANTS[1,2,6], Ben VAN CALSTER[1,2,4,*]

**Affiliations:**

[1]Department of Development and Regeneration, KU Leuven, Leuven, Belgium.

[2]Leuven Unit for Health Technology Assessment Research (LUHTAR), KU Leuven, Leuven, Belgium

[3]Department of Biomedical Data Sciences, Leiden University Medical Centre, Leiden, the Netherlands.

[4]Julius Centre for Health Sciences and Primary Care, University Medical Centre Utrecht, Utrecht, Netherlands

[5]Dept Obstetrics & Gynecology, University Hospitals Leuven, Leuven, Belgium.

[6]Department of Epidemiology, CAPHRI Care and Public Health Research Institute, Maastricht University, Maastricht, Netherlands.

*Corresponding author: Ben Van Calster

Email: B.J.vanCalster@umcutrecht.nl

Herestraat 49 box 805 | B3000 Leuven | Belgium

## Word count: 2999

## Abstract

**Background and Objective:** Clinical prediction models are commonly evaluated regarding performance for a population, although decisions are made for individuals. The classic view relates uncertainty in risk estimates for individuals to sample size (estimation uncertainty) while other sources are model uncertainty (variability in modeling choices) and applicability uncertainty (variability in measurement procedures and between populations). We aim to illustrate the uncertainty of prediction models in estimating individual risks with an ovarian cancer example.

**Methods:** We used real and synthetic data for ovarian cancer diagnosis to train 59400 models with variations in estimation, model, and applicability uncertainty. We then used these models to estimate the probability of ovarian cancer in a fixed test set of 100 patients and evaluate the variability in individual estimates.

**Results:** We show empirically that estimation uncertainty can be strongly dominated by model uncertainty and applicability uncertainty, even for models that perform well at the population level. Estimation uncertainty decreased considerably with increasing training sample size, whereas model and applicability uncertainty remained large.

**Conclusion:** Individual risk estimates are far more uncertain than often assumed. Model uncertainty and applicability uncertainty usually remain invisible when prediction models or algorithms are based on a single study. Predictive algorithms should inform, not dictate, care and support personalization through clinician-patient interaction.

**What is new?**

- **Key findings:** Individual AI risk predictions carry massive uncertainty driven by varying model choices and differences between patient populations—a problem that simply using larger datasets cannot fix.
- **What this adds to what was known:** A new framework is introduced to understand uncertainty, showing that traditional evaluations only consider sample size and hide important other sources of uncertainty.
- **What is the implication and what should change now?** Health AI should only inform, not dictate, clinical decisions. True personalized medicine must rely on human interaction and shared decision-making between the clinician and patient, rather than blind trust in AI outputs.

## Introduction

Clinical risk prediction models, which are increasingly based on artificial intelligence (AI), estimate the probability of a medical event occurring in a patient, conditional on a set of predictors.[1] Examples include models to estimate the 10-year risk of coronary heart disease [2], early mortality in cardiac surgical patients [3], the risk of ovarian cancer [4], the 10-year mortality risk of breast cancer patients [5], or the risk of deterioration in hospitalized patients with COVID.[6]

The performance of prediction models is usually assessed at the population level. To make informed decisions for an individual patient, however, the focus is on the risk estimates for that individual. The trustworthiness of individual risk estimates is commonly ignored in practice, and rarely evaluated or quantified.[8] We aim to present a structured framework of the key sources of uncertainty in risk estimation for individuals, and to expose the magnitude of uncertainty arising from sources that are typically hidden in practice. We illustrate this with a case study on estimating the risk of cancer in patients presenting with an ovarian tumor.

## A framework for uncertainty

Uncertainty and the concept of 'individual risk' have been studied for centuries.[7–16] An individuals' risk of an event as estimated by a prediction model refers to the probability of that event in the subgroup of individuals with the same predictor values. Truly individual risks do not exist because of the 'reference class problem': an individual can belong to an infinite number of subgroups. By choosing a different set of predictors to condition on, the reference class changes, and so does the risk. Yet, despite the philosophical challenges in defining a true individual risk, risk estimates are routinely used in practice, making it crucial to understand the different types of uncertainty they carry. For predictive modeling, a basic decomposition of uncertainty is into aleatoric and epistemic uncertainty (**Table 1 and Figure 1**).[17,18] Aleatoric uncertainty refers to the inherent random nature of the event we want to predict. Two patients who have the same values for the predictors in the best possible model may still have a different outcome. This does not invalidate the prediction. In contrast, epistemic uncertainty refers to lack of knowledge about the best possible model. We break down epistemic uncertainty for predictive modeling into three key categories: estimation, model, and applicability uncertainty (**Table 1**).

Estimation uncertainty (sometimes referred to as approximation uncertainty) refers to variability in the parameter estimates of a fitted model.[19,20] Given a specified modeling approach, using different training samples of the same size and from the same population leads to different parameter estimates. This variability decreases with increasing sample size. The estimation uncertainty of risk estimates can be quantified using classic statistical confidence intervals, for example.

Model uncertainty refers to uncertainty about model specification.[17,21,22] Specifying a model involves a substantial number of critical choices. This includes selecting the model class, the candidate predictors, and the methods for model configuration (e.g., hyperparameter tuning). Applicability uncertainty relates to the use of the model in a specific setting and context. We distinguish between two dimensions of applicability uncertainty.[21] Data uncertainty relates to variability in the measurement of predictors and the outcome. Predictors and outcomes can have different definitions, measurement procedures, types of measurement error, and missing data patterns. Population uncertainty refers to the general phenomenon of data drift: differences in populations and clinical practices between hospitals, countries, or time periods.[23]

Within epistemic uncertainty, the researcher is a transversal source of uncertainty. Researchers exhibit variability in their levels of subject knowledge regarding the data, modeling expertise, and established best practices in predictive analytics. Systematic reviews reveal high frequencies of practices that are generally discouraged.[24–27] Examples include unnecessary dichotomization of predictors, unjustified discarding of incomplete cases, insufficient protection against overfit, using default values for hyperparameters, or ignoring competing events when predicting time-to-event outcomes. Additionally, researchers often have differing opinions and preferences regarding data collection and modeling approaches, such as the choice between regression, boosting, or deep learning algorithms. This variability in researcher's knowledge, practices, and methodological choices constitutes a significant and transversal source of uncertainty.[19,20,28–31]

## Illustrative case study: Estimating the risk of malignant ovarian cancer

Suppose we want to develop a model to estimate the risk that a suspicious ovarian tumor is malignant, to guide the decision to refer for specialized surgery by a gynecological oncologist. We use a dataset including 1122 patients from the University Hospitals Leuven (Belgium) who had surgery between 1999 and 2015.[4,32] One fixed test set of 100 patients was set aside to quantify variability in the risk estimates. Using the remaining 1022 patients, a synthetic training dataset of 100,000 patients is generated to explore the effect of sample size on uncertainty (**see Figures S1 and S2 and Appendix 1 in Supplementary materials and methods** for details).

Our preferred modeling approach is logistic regression using the following six predictors: patient age in years, maximal diameter of the lesion in mm, proportion solid tissue (values between 0 and 1), CA125 (IU/L), bilaterality (whether or not there were tumors on both ovaries), and presence of papillary projections with blood flow. The ultrasound examination was performed according to the standardized terms and definitions of the International Ovarian Tumor Analysis (IOTA) consortium.[33] The four continuous predictors are modeled using restricted cubic splines with three knots. The minimum required sample size for such a model (10 parameters, 37% prevalence of malignancy, and an assumed AUROC of 0.88 based on the literature) is 359.[34,35] CA125 had 31% missing values, which are related to the ultrasound appearance. Reasonably assuming 'missingness at random' (MAR) we used regression imputation based on the other five predictors.[36]

We applied the following steps to train the model: (1) we randomly drew a training sample of 400 patients with replacement, (2) imputed CA125 in the training sample, (3) trained the model, (4) applied the imputation model to the test set, and (5) applied the prediction model to the test set.

### Variations on the modeling strategy

To evaluate the impact of all uncertainty categories, we considered simple variations on the preferred modeling approach (**Supplementary Table S1**). Regarding model uncertainty, we used 33 variations based on different approaches for handling continuous variables options, variable selection, penalization, and by replacing logistic regression with random forests or xgboost with various options for the depth of the trees. Regarding data uncertainty, we used 6 variations based on different approaches for defining tumor size and proportion solid tissue and for handling missing data. Regarding population uncertainty, we used 3 variations based on the hospital from which training data were recruited (Leuven, Rome, Malmö) (See **Supplementary Table S2** for center-specific detailed information). Taken together, we have 594 variations or model training scenarios. Regarding estimation uncertainty, we repeated each of the 594 scenarios using 100 randomly drawn training samples, resulting in 59,400 trained models; then we repeated everything for training sample sizes of 2,000 and 10,000. The test set is always the same set of 100 real patients from Leuven. Hence, for each sample size, we obtain 59,400 probability estimates for each test set patient.

For comparison we repeated the exercise with training sample size 400 using real data (1022 patients from Leuven, 1131 from Rome, 1048 from Malmö.

## Model performance on the test set

We evaluated population-level performance using the AUROC for discrimination, estimated calibration index (ECI) for calibration (ECI=0 means perfect moderate calibration), and relative utility (RU) for clinical utility (RU≤0 means no utility) [37,38]. To evaluate individual-level variability, we calculated for each patient the 95% range and decision uncertainty (DU) of 59,400 estimated risks per sample size. The 95% range is the estimated risk at the 97.5th percentile minus the estimated risk at the 2.5th percentile. DU is lowest of the proportions of models that yield an estimated risk <0.1 vs ≥0.1. This is a common threshold in the setting of ovarian cancer diagnosis.[39] If all models yield a risk on the same side of the threshold, there is no decision uncertainty (DU=0). If half of the models yield a risk above the threshold, and half below, there is maximal uncertainty (DU=0.5). See **Appendix 2 in Supplementary materials and methods** for details. Materials and instructions to reproduce the methodology and results are provided in the OSF repository (https://osf.io/uxvhq/).

# Results

## Estimation uncertainty

Consider the estimated risk of ovarian cancer for 100 new patients (i.e. patients whose data was not used in model building) according to 100 identical models, each fitted on a different random sample from one population. For some patients, the risk estimates (on a scale from 0 to 1) varied considerably, particularly at small training sample sizes **(Figure 2, left**). The mean 95% range of estimated risks for a new patient decreased from 0.22 for training sets with N=400 to 0.04 for training sets with N=10,000 (**Table 2).** When a risk estimate of 0.1 or higher was used as the threshold to recommend surgery for a single new patient, on average 6% of models trained on a dataset of 400 patients would suggest a different decision than the majority of models. Increasing the training set to 10,000 patients reduced this decision uncertainty to 1%. We observed that estimation uncertainty is higher and decreases more slowly with increasing training sample size for tree-based models compared to logistic regression (**Table S3, Supplementary Figure S3**). In contrast to the indicators of uncertainty at the individual level, the population level discrimination and calibration remained rather stable across training sample sizes. The positive population level RU indicates the model is useful to support clinical decision-making at all sample sizes.

## Model and applicability uncertainty

When also acknowledging model and applicability uncertainty, the differences in estimated ovarian cancer risks for any single patient from a total of 59,400 models were considerably larger than when acknowledging estimation uncertainty alone (**Figure 2, right**). Even when the training sample size was 10,000, the 95% range in risk estimates for a single patient was on average 0.39 (range 0.05 to 0.99). The highest 95% risk range for a single patient was 0.99, nearly covering all possible risk estimates between 0 and 1.

In contrast to approximation uncertainty, increasing the training sample size led to minimal reductions in model uncertainty and applicability uncertainty. On average, 12% of models trained on 400 patients would lead to a different clinical decision for a single patient than most models suggest. When training was done on data of 10,000 patients, the decision uncertainty was still 11%. Note that this is substantial compared to the maximum decision uncertainty of 50% (when 50% of models suggest operating and 50% of models suggest not operating). Findings were similar for any class of models **(Supplementary Table S3**).

The mean AUROC (0.92-0.93) and ECI (0.09-0.11) values were similar to the values observed when we only considered estimation uncertainty (**Table 2**).Detailed results about estimation, model and applicability uncertainty available in **Supplementary materials and methods** (Synthetic data: **Supplementary Figures S3-S6; real data: Supplementary Figures S7-S10 and Supplementary Tables S4-S5**).

## Discussion

Precision medicine aims to target interventions towards the specific characteristics of the individual. In this context, prediction models and AI algorithms can be valuable to improve decision making. Our study shows that, even though prediction models may perform very well at the population level (high discrimination, good calibration and clinical utility to support decisions), the uncertainty in estimated risks and suggested decisions for a single patient is often uncomfortably large.

Calibration assesses whether estimated risks correspond to observed proportions for (sub)groups of patients, typically using calibration metrics and plots, but does not inform on the uncertainty of risk estimates for an individual patient. The classic approach towards uncertainty assessment for an individual involves confidence intervals or other quantifications of variability that assume that modeling, data, and population decisions are fixed (estimation uncertainty, **Table 1**).[19,40,41] Recent research on clinical prediction models has focused on estimation uncertainty, i.e. instability as a function of sample size [19,20,31]. We demonstrated that the magnitude of model uncertainty and applicability uncertainty may extend far beyond estimation uncertainty. Estimation uncertainty differs from other sources of epistemic uncertainty in two aspects. First, it reduces more strongly with increasing sample size than other sources of uncertainty. Second, it can be visualized and quantified, whereas other sources are mostly invisible and hardly quantifiable. A few studies highlighted that modeling choices including model class, predictors, and missing data imputation techniques may impact on individual risk estimates.[28–30] In the current analyses, we quantified how data-related choices (in measurement, definition and populations) further impact uncertainty. The number of decisions that are taken to develop a model is potentially infinite, yet we usually only see one specific model.

It may seem paradoxical that using prediction models for decision making can benefit outcomes at a population level without having trustworthy risk estimates for individuals from that population. This may cause skepticism about the acceptability of data-driven clinical decision support tools, either based on classical regression approaches or novel AI algorithms. We argue that such skepticism is not needed. Clinical utility at the population level, for example in terms of cost-effectiveness, is sufficient to warrant the use of a decision strategy based on a model. We may need to accept that two models may have the same clinical utility yet very different risk estimates for the same individual, often to the extent that model A recommends treatment initiation, but model B does not. The enormous uncertainty on the individual level does have consequences. First, the uncertainty should be acknowledged and considered when discussing risks and treatment options with patients. Given the considerable impact of poorly quantifiable model and applicability uncertainty, further research is needed on how to integrate model predictions in the shared decision making process. Second, while we should strive to reduce epistemic uncertainty in clinical prediction models, we must also recognize and accept it as an unavoidable element of medical decision-making. Finally, we should be humble regarding the promises of personalized medicine.[42] Health AI will always provide highly uncertain group level estimations based on their training data, even if the group is defined very narrowly. As a consequence, the estimated risk for the individual is not personalized, yet the shared decision making based on this estimation can be.

Researchers have a variety of tools at their disposal to reduce or embrace epistemic uncertainty, listed in **Table 1.** First, to reduce estimation uncertainty, we need to develop models on sufficiently large sample sizes.[43] Whereas there are elegant sample size calculation procedures for prediction models based on regression, such procedures are lacking for machine learning algorithms, which may generally require large sample size to allow for flexible modeling.[44,45] Further, methods that allow prediction abstention due to uncertainty have been proposed.[41,46] Even though not all sources of uncertainty are included, it may be a good start to abstain from predictions based merely on estimation uncertainty, for example requiring at least an effective sample size of 10 patients per prediction.[31,40] Second, a better agreement and education about good prediction modeling practices could reduce model uncertainty, including the effects of modeler uncertainty, to some degree. Third, to reduce data uncertainty, the use of prospective study designs can help. Retrospective studies tend to have increased uncertainty regarding the definition, measurement procedure, and availability of relevant predictors and outcome variables. Prospective studies, including randomized controlled trials, typically allow for better standardization of definitions and measurement protocols.[47] For example, a set of 'terms and definitions' has become standard in the field of ovarian tumors [33] and 'common data elements' have been defined for traumatic brain injury studies.[48,49] While sample size is important for model development, data quality should be prioritized over quantity for clinical prediction model development. Although a huge dataset further reduces estimation uncertainty, it may increase data uncertainty due to data quality problems. Fourth, population uncertainty may be exposed by assessing population differences and performance heterogeneity across sites in multicenter development and validation studies.[50] Monitoring and updating of prediction models can improve local performance and further decrease uncertainty. Ideally, this includes assessing the causes of the heterogeneity and performance changes over time.[51,52]

The strength of our demonstration on ovarian cancer diagnostic modeling is the extensive investigation of different types of uncertainty on empirical data, and its presentation in easy-to-interpret metrics and simple visualizations. At the same time, this approach has a few limitations. First, because the primary objective was to illustrate the uncertainty in risk prediction using real-world retrospective data, this study was conducted without a formal, pre-specified protocol. The retrospective nature constrained the available data in our choices related to data and population uncertainty, likely leading to an underestimation of applicability uncertainty. Second, to investigate the results for models built on large training datasets, we created a larger synthetic dataset with identical properties as the original dataset. Third, the event of interest in our dataset was relatively common (37% of tumors were malignant), which yields larger values for the 95% range compared to situations where the event is rare. Of note, smaller 95% ranges in cases with rare events may still affect decision uncertainty, if the decision threshold falls around instable predictions. Fourth, diagnostic models for ovarian malignancy have high AUROC values (above 0.9). While higher levels of uncertainty might be present for prediction problems with lower discrimination, the amount of individual risk uncertainty is striking for a model that is well able to distinguish between patients with the event and patients without it at the population level. Finally, although we considered a large number (594) of scenarios, more scenarios could have been conceived. For example, we did not vary the set of six predictors, and our variations in the tree-based algorithms were less extensive than those for the logistic regression analysis. These characteristics of the data and selected scenarios enforce the conclusions of our study, given that most of these issues have most likely underestimated the full uncertainty.

In conclusion, a fundamental problem in risk prediction is that individual risk estimates may be highly uncertain, even for models with high discriminatory ability that improve outcomes for the population when used to guide decision making. The uncertainty for predictions at the individual level will often be considerably larger than suggested by classic uncertainty measures such as 95% confidence intervals or novel measures such as effective sample size.[40] This uncertainty does not invalidate the clinical utility of well-developed prediction models at a population level, but urges great caution when interpreting and discussing risk estimates with individual patients in clinical practice.

**Acknowledgments:** Acknowledgments follow the references and notes list but are not numbered. Start with text that acknowledges non-author contributions (including disclosure of any editing services or AI-assisted technologies used in preparation of the manuscript), then complete each of the sections below as separate paragraphs.

**Funding:** This research is supported by Research Foundation – Flanders (FWO) grants G097322N, G049312N, G0B4716N, and 12F3114N to BVC and/or DTi, KU Leuven internal grants C24M/20/064 and C24/15/037 to BVC and/or DTi, ZoNMW VIDI grant 09150172310023 to LW. DTi is a senior clinical investigator of FWO. No funding source had any role in the study design, data collection,data analysis, data interpretation, writing, or submission of this report.

**Author contributions:** Conceptualization: BVC. Methodology: LB, ES, LW, DTh, BVC. Investigation: LB, BVC. Visualization: LB, BVC. Funding acquisition: DTi, LW, BVC. Project administration: LB. Supervision: ES, BVC. Writing – original draft: LB, LW, BVC. Writing – review & editing: LB, ES, DTi, DTh, LW, BVC.

**Competing interests:** Authors declare that they have no competing interests.

**Data and materials availability:** Synthetic and real data are not publicly available, but we provide the individual risk estimations of all the different models and explanatory code to reproduce all the results in this work in the OSF repository (https://osf.io/uxvhq/)

**Declaration of generative AI and AI-assisted technologies in the manuscript preparation process:** During the preparation of this work the authors used Google Gemini Pro in order to proofread the text and improve code readability. After using this tool/service, the authors reviewed and edited the content as needed and take full responsibility for the content of the published article.

**Table 1**. Overview of the categories of epistemic uncertainty.

| Description | Examples | Actions |
|---|---|---|
| **ESTIMATION UNCERTAINTY** | | |
| Instability of the fitted model due to having finite training sample size. | - If we collect two datasets from the same population with the same sample size and sampling procedures, and train the same model on both datasets, the estimated model parameters will be different. | - Increase sample size.<br>- Quantify instability, e.g. with 95% confidence intervals, or 'effective sample size' [40] associated with a prediction. |
| **MODEL UNCERTAINTY** | | |
| Uncertainty due to lack of knowledge about the optimal model specification. | - Choosing the model class (e.g. regression, random forest, convolutional neural network).<br>- Specifying features, feature transformations, and feature selection strategy.<br>- Selecting which hyperparameters to tune and the tuning strategy. | - Improve education about good modeling practice.<br>- Conduct literature searches or exploratory studies to identify features.<br>- Conduct pilot studies to compare model classes. |
| **APPLICABILITY UNCERTAINTY REGARDING DATA** | | |
| Uncertainty due to variability in data collection procedures. | - Variables may be defined in different ways. This applies to predictor and outcome measurements (e.g. central-line associated bloodstream infections [53] and delirium).<br>- Measurement procedures vary. Biomarkers can be measured with different assay kits using various body fluids (e.g., serum or urine). Patient characteristics or behaviors can be self-reported or measured objectively. Medical images can be taken using different types of machines with different acquisition settings. Moreover, inter-rater variability occurs when manually labeling medical images [54].<br>- The fraction, cause and handling of missing data may vary. In electronic health records data, the decision to acquire a certain measurement varies between hospital or even physician[55]. In prospective observational studies, there is variability regarding the effort to have complete measurements, local clinical protocols, and the financial cost of measurements. Missing values can be ignored ('complete case' analysis) or addressed using various statistical approaches (such as 'imputation'). | - Standardize definitions and measurement procedures.<br>- Conduct studies to empirically identify the best measurement procedure.<br>- Secure resources to ensure proper data collection.<br>- Avoid retrospective studies.<br>- For prospective studies, invest in real-time monitoring of data collection.<br>- If retrospective studies are unavoidable, invest in proper data extraction.<br>- Invest in cleaning, linkage, and harmonization in line with the specific research question and intended application. |
| **APPLICABILITY UNCERTAINTY REGARDING POPULATIONS** | | |
| Uncertainty due to variability between population characteristics in different locations. | - Populations may vary in case-mix (distribution of predictors and variables not included as predictors), prevalence or incidence of the outcome event of interest, treatment pathways for prognostic outcomes, and/or in the associations between predictors and the outcome.<br>- Within a specific setting, there is population uncertainty due to phenomena such as population drift: populations tend to evolve over time, owing to evolutions in care or referral patterns [56].<br>- When a model is developed, the specific inclusion and exclusion criteria may further increase population uncertainty. | - Conduct multicenter studies to develop and validate models.<br>- Quantify heterogeneity using methods such as leave-center-out cross-validation and meta-analysis of center-specific performance with prediction intervals.<br>- Consider local model validation and updating.<br>- Investigate causes of heterogeneity to allow targeted model improvements [52]. |

**Table 2**. Population level performance and individual risk uncertainty for the 100 test set patients.

| | | | Population level performance | | | Individual level uncertainty | |
|---|---|---|---|---|---|---|---|
| Source of uncertainty | Training data | # models | AUROC, mean (range) | ECI, mean (range) | RU, median[a] (range) | 95% range, mean (range) | DU, mean (range) |
| Estimation uncertainty | N=400 | 100 | .92 (.89; .93) | .11 (.02; .21) | .05 (-.61; .49) | .22 (.01; .65) | .06 (0; .49) |
| | N= 2000 | 100 | .92 (.91; .93) | .11 (.06; .14) | .20 (-.02; .49) | .10 (.01; .36) | .02 (0; .33) |
| | N=10000 | 100 | .93 (.92; .93) | .11 (.10; .13) | .18 (.02; .37) | .04 (.00; .14) | .01 (0; .31) |
| All sources of uncertainty | N=400 | 59400 | .92 (.79; .98) | .11 (<.01; .93) | .16 (-2.2; .82) | .46 (.07; .91) | .12 (0; .50) |
| | N=2000 | 59400 | .93 (.84; .98) | .10 (<.01; .68) | .18 (-2.0; .82) | .41 (.06; .95) | .11 (0; .50) |
| | N=10000 | 59400 | .93 (.83; .98) | .09 (<.01; .55) | .18 (-1.5; .84) | .39 (.06; .98) | .11 (0; .50) |

AUROC: Area under the receiver operating characteristic curve; AUROC 0.5 indicates that the model cannot discriminate between patients with and without ovarian cancer, AUROC 1 indicates perfect discrimination. ECI: Estimated calibration index; ECI is scaled between 0 and 1 with 0 indicating that estimated risks correctly estimate risk at the population level. RU: Relative utility; RU an take on values between $-\infty$ and 1 where positive values indicate the model has clinical utility to support decision making and values ≤ 0 indicate that the model has no clinical utility. 95% range: range of the middle 95% of risk estimates across all models for one individual; 0 indicates no uncertainty, 1 indicates maximal uncertainty. DU: Decision uncertainty for an individual when using the model at a threshold of 0.1 on the risk of malignancy to suggest a management decision. DU ranges between 0 and 0.5, with 0 indicating that all models suggest the same decision, and 0.5 that half of the models suggest to operate and the other half suggest conservative management.

[a] We used median because negative values for RU can be very low and distort the results.

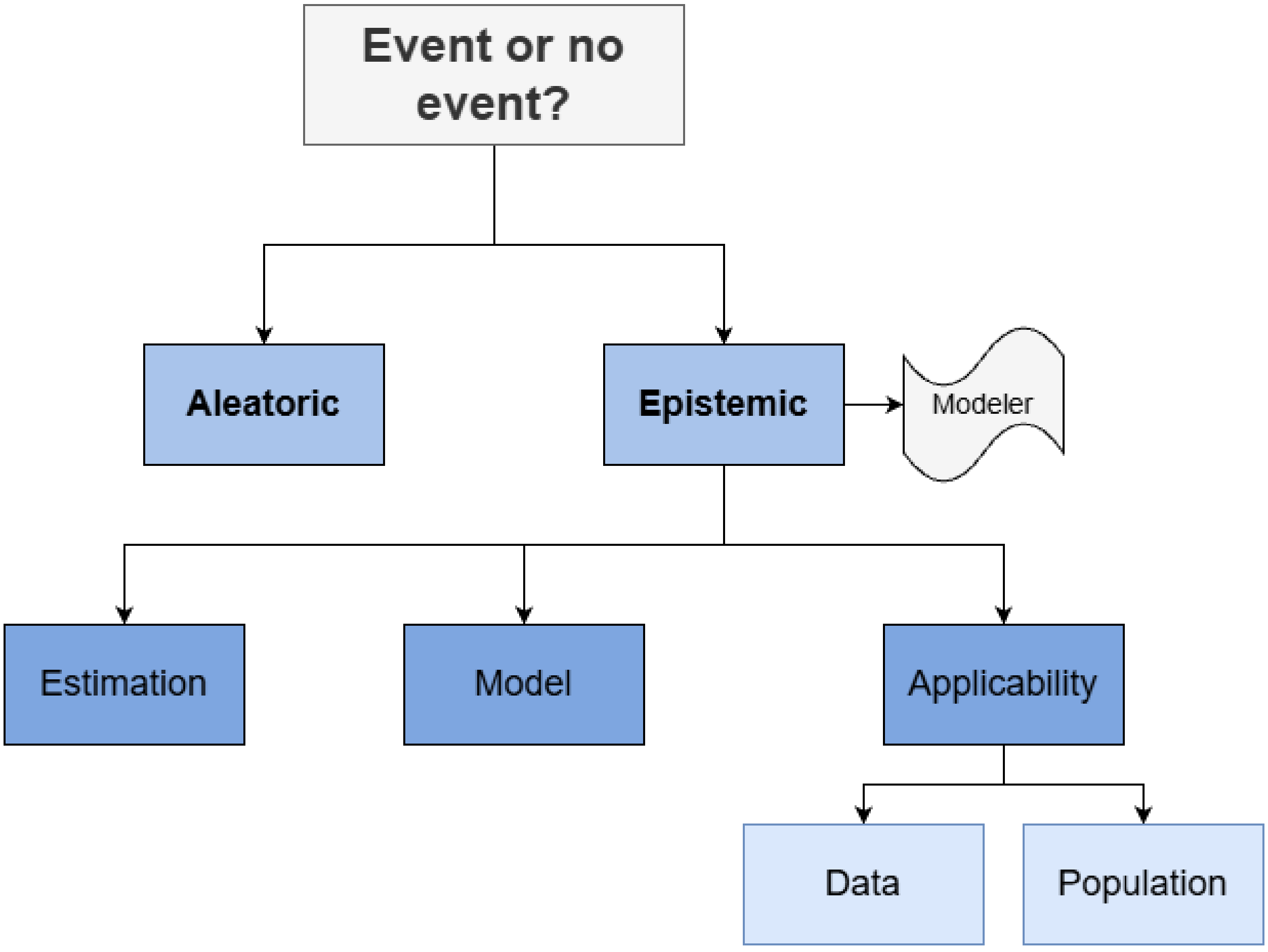

**Fig. 1.** A framework of uncertainty when predicting risk for the individuals.

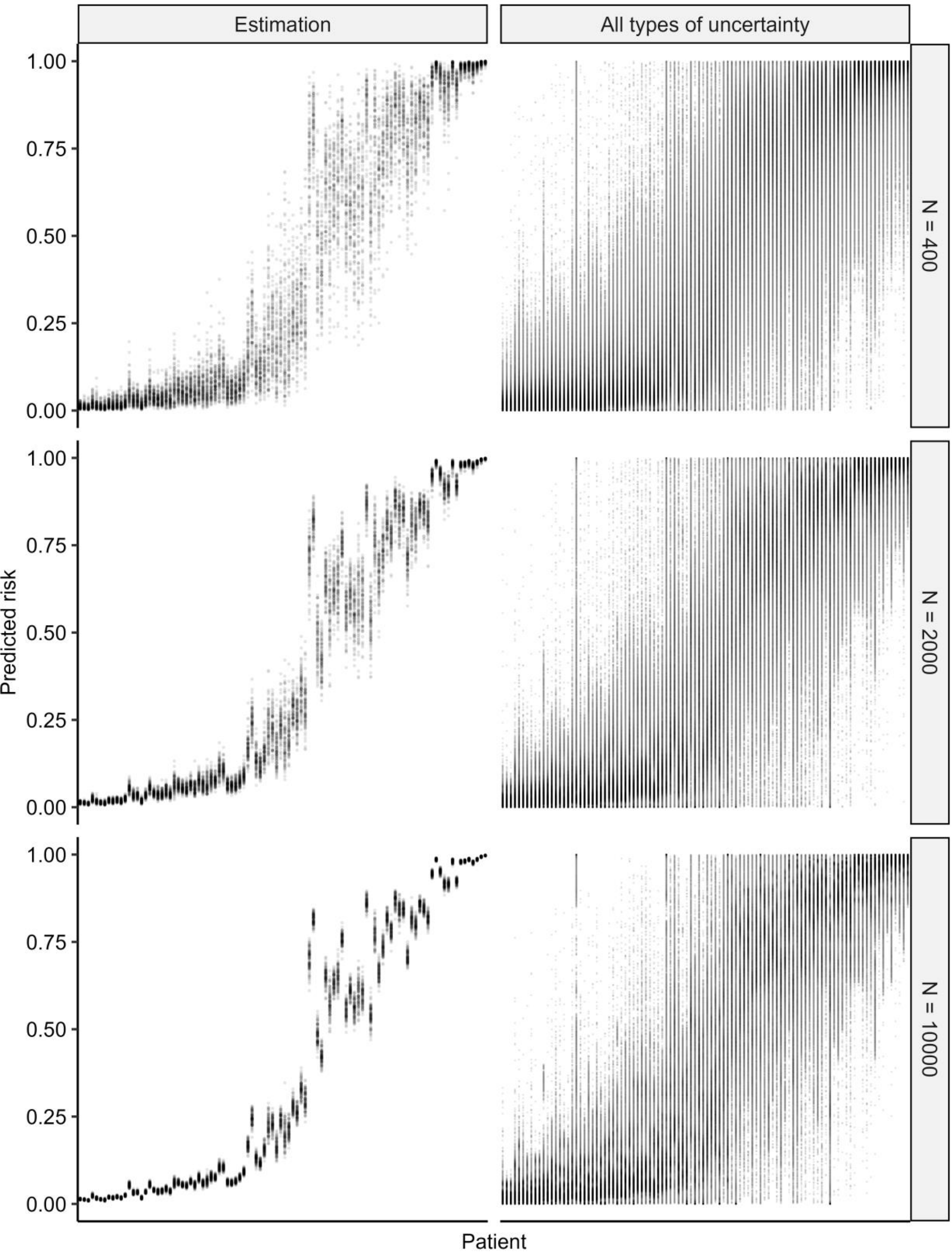

**Fig. 2. Visualization of the uncertainty in the estimated risks for individual patients.** Panels on the left show estimation uncertainty for the main model based on training sample sizes of 400 (top), 2000 (middle) and 10000 (bottom). Each panel contains the individual estimated risks (y-axis) for 100 test patients (x-axis) by 100 models based on randomly selected training datasets. The main model used unpenalized logistic regression with restricted cubic splines, with regression imputation for missing CA125, no variable selection, using training data from Leuven and with lesion size measured using the maximal diameter. Panels on the right show the total uncertainty for all 59,400 models. Each panel contains the individual estimated risks (y-axis) for 100 patients (x-axis) by 59,400 models based on combinations of modeling strategies, data properties, populations included at training (**Supplementary Table S1**) and on randomly selected training datasets. This means that each of the 100 patients has 59,400 predicted estimated risks.

# Supplementary materials and methods for

## The fundamental problem of risk prediction for individuals: health AI, uncertainty, and personalized medicine

Lasai BARREÑADA, Ewout W STEYERBERG, Dirk TIMMERMAN, Doranne THOMASSEN, Laure WYNANTS, Ben VAN CALSTER
Corresponding author: ben.vancalster@kuleuven.be

**Appendix 1. Method for generating synthetic data**

We generated synthetic data using the synthpop package in R (https://synthpop.org.uk/). This package uses true data to generate a data generating model that will mimic the distribution of each variable in the true dataset and their correlation. Then with this data generating model we can generate as much synthetic observations as desired preserving the relationship between the variables but removing all sensitive individual patient data and allowing to generate a larger sample than the original one (see **Table S1** for variable distribution in the new generated datasets) [1]. In our analysis, we first select a fixed test set (same as in the previous analysis) from the true population. Then with the rest of the dataset in each center, we generate 100.000 observations per center using *synthpop*. We generate synthetic data using all predictors (both measures of diameter and volumes), with log CA125 and outcome. The method is able to synthetically generate missing values that will then be imputed according to the modeling approach. The synthetic data and the true data are not publicly available but the code to generate it and the estimated probabilities by the different models are available in the OSF repository (osf.io/uxvhq).

We measure the quality or utility of synthetic data based on the similarity between the true data distribution and synthetic data distributions, correlation of predictors and with the results obtained from fitting a prediction model with both datasets. To explore this we use histograms, correlations and scatterplots of 2 variables.

Graphical evaluation of univariate and bivariate predictors is presented in **Figures S1-S2**. The bivariate and univariate distribution are very similar which means that the synthetic data is correctly being generated. The only variables that present differences in distributions and correlations are the semi-continuous proportion of solid tissue which tends to not mimic perfectly the true data. However, this seems to have not an important impact on the models.

We fit the prediction model for the main analysis (logistic model) after imputing CA125 with regression imputation in the whole individual patient dataset of Leuven and the synthetic dataset of Leuven. We obtain these coefficients for the model with RCS and without RCS:

| Variable | True model with RCS | Synthetic model with RCS |
|---|---|---|
| Intercept | -10.62 | -10.52 |
| Age | 0.05 | 0.04 |
| Age' | -0.02 | 0.00 |
| lesdmax | 0.71 | 1.10 |
| lesdmax' | 0.56 | 0.01 |
| propsol | 4.64 | 3.92 |
| propsol' | -3.73 | -2.74 |
| bilateral | 0.70 | 0.62 |
| papflow | 2.05 | 2.22 |
| lCA125i | 0.64 | 0.30 |
| lCA125i' | 0.10 | 0.86 |

| Variable | True model | Synthetic model |
|---|---|---|
| Intercept | -11.70 | -11.20 |
| Age | 0.03 | 0.04 |
| lesdmax | 1.16 | 1.05 |
| propsol | 2.26 | 2.18 |
| bilateral | 0.67 | 0.61 |
| papflow | 2.41 | 2.47 |
| lCA125i | 0.70 | 0.67 |

After conducting the analysis, we noticed that the results of **Table S3** with sample size 400 are not equivalent to the results of **Table S5.** This is probably due to an important underestimation of the total uncertainty in the IP data analysis caused by the similarity between the randomly selected training samples. We checked this by repeating the analysis with synthetic data of the same size as the patient data, therefore generating 1022,1053,1141 synthetic patients from Leuven, Malmö and Rome respectively and the results were identical to **Table S5**.

**Fig. S1. Relative frequency histogram of synthetic and observed data in the variables used to generate the synthetic data.**

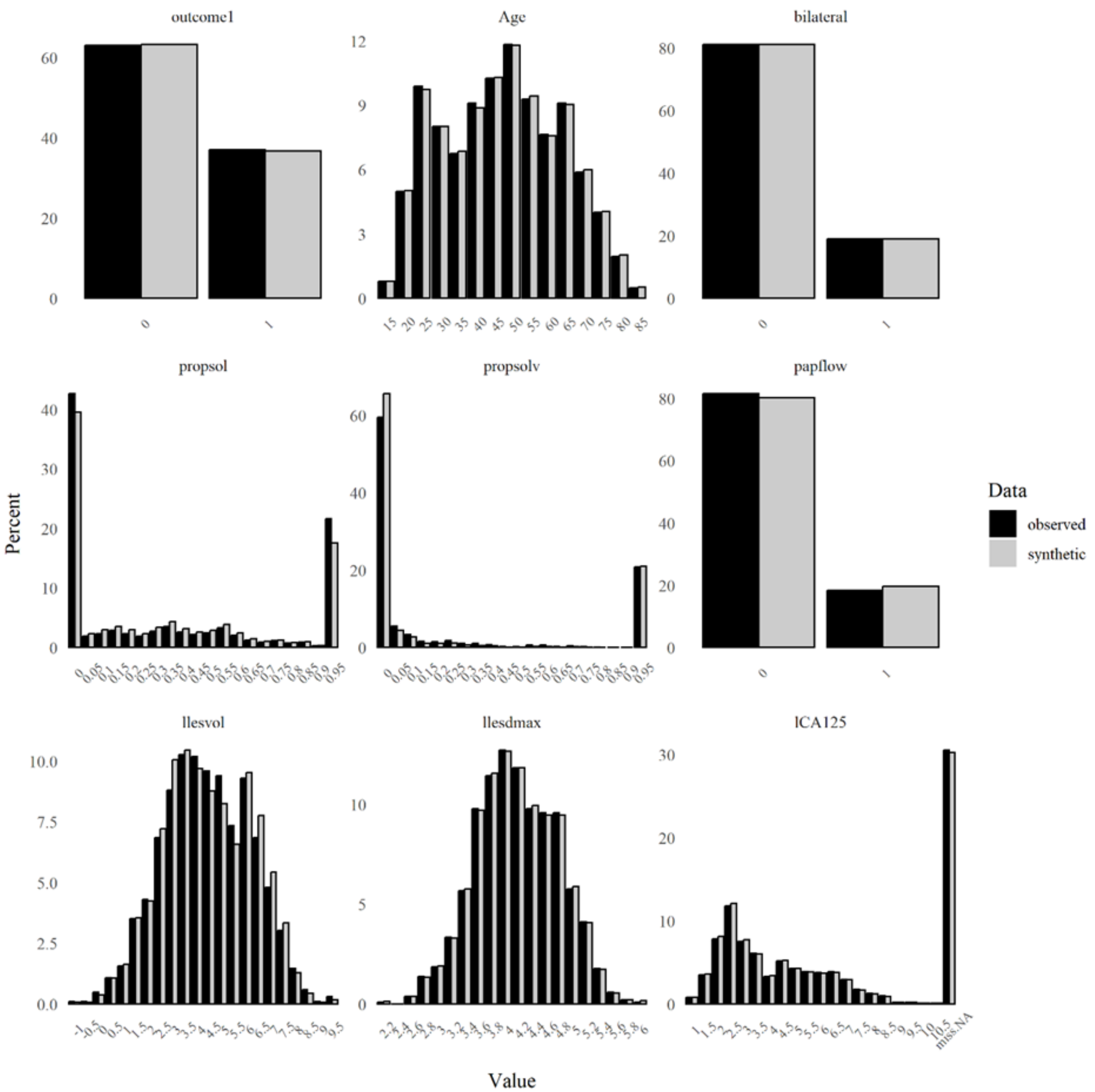

**Fig. S2. Two by two comparison of continuous predictors in synthetic and true data. Upper diagonal plots show Spearman correlations.**

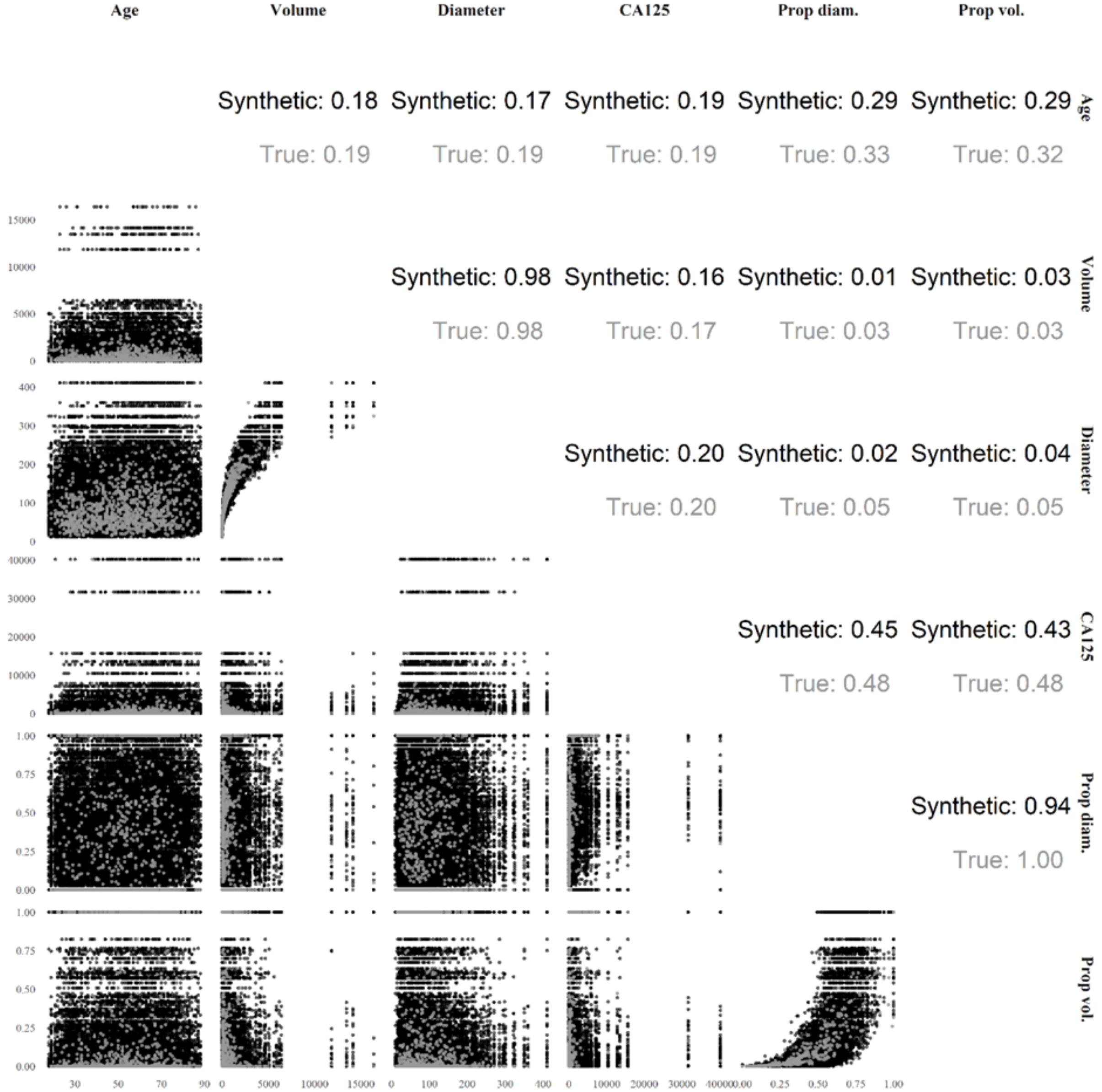

**Appendix 2. Metrics explained**

We evaluate model performance and consistency across a fixed test set of N = 100 patients, indexed by i (where i = 1, …, N). Each patient i receives J risk estimates, denoted as $p_{ij}$, produced by J different models (indexed by j = 1, …, J).

**95% Risk Range (95R)**

The 95% risk range (denoted $95R_{ij}$ for patient *i* in J models captures the spread of model-predicted risks, focusing on the central 95% of values. It is calculated as:

$$95R_{ij} = F_N^{-1}(0.975) - F_N^{-1}(0.025)$$

where $F_N^{-1}$ is the inverse of the empirical distribution function of the J risk predictions for ith patient $F_N(\mathbf{p_{ij.}})$. This range reflects the level of uncertainty for that patient across the different models.

**Decision uncertainty (DU)**

Decision uncertainty quantifies how often the models disagree on the treatment decision for a patient based on a specified risk threshold. For each patient i, we compute the proportion of models that predict a risk on the opposite side of the decision threshold from the majority

$$DU_{ij} = \frac{\min\left(\sum_{j=1}^{J} p_{ij} \geq threshold\,, \sum_{j=1}^{J} p_{ij} < threshold\right)}{J}$$

The overall decision uncertainty (DU) across all patients for the J subset of models is then the average of these values:

$$DU_j = \sum_{i=1}^{N} DU_{ij}$$

Note that J varies according to the type of uncertainty analysis, but N always denotes the 100 test patients. For instance, in the case of estimation uncertainty, N is 100 and J is 100. If for a patient 30/100 models predict a risk higher than the decision threshold that patients' decision uncertainty is 0.3. Averaging over all patients will give the DU of that group of models. DU is bounded between 0 and 0.5, with 0 denoting perfect concordance between all models and 0.5 maximal discordance.

**Net benefit and relative utility**

Net Benefit (NB), Net Benefit of treating all patients ($NB_{TA)}$, and Relative Utility ($RU$) are defined as follows:

$$NB_j = \frac{TP_j}{N} - \frac{FP_j}{N} \times \frac{threshold}{1 - threshold}$$

$$NB_{TA} = \frac{49}{100} - \frac{51}{100} \times \frac{threshold}{1 - threshold}$$

$$RU_j = \frac{NB_j - \max(0, NB_{TA})}{0.49 - \max(0, NB_{TA})}$$

with $TP$ true positives, $FP$ false positives, $threshold$ the selected risk threshold and N the number of patients evaluated. In our case of ovarian cancer diagnosis, a common and clinically accepted threshold is 0.1.

**Fig. S3. Scatterplot of individual risk by patient to illustrate estimation uncertainty in synthetic data by type of model and training sample size. Each patient has 100 individual risk generated by models where the only variation was training sample.**

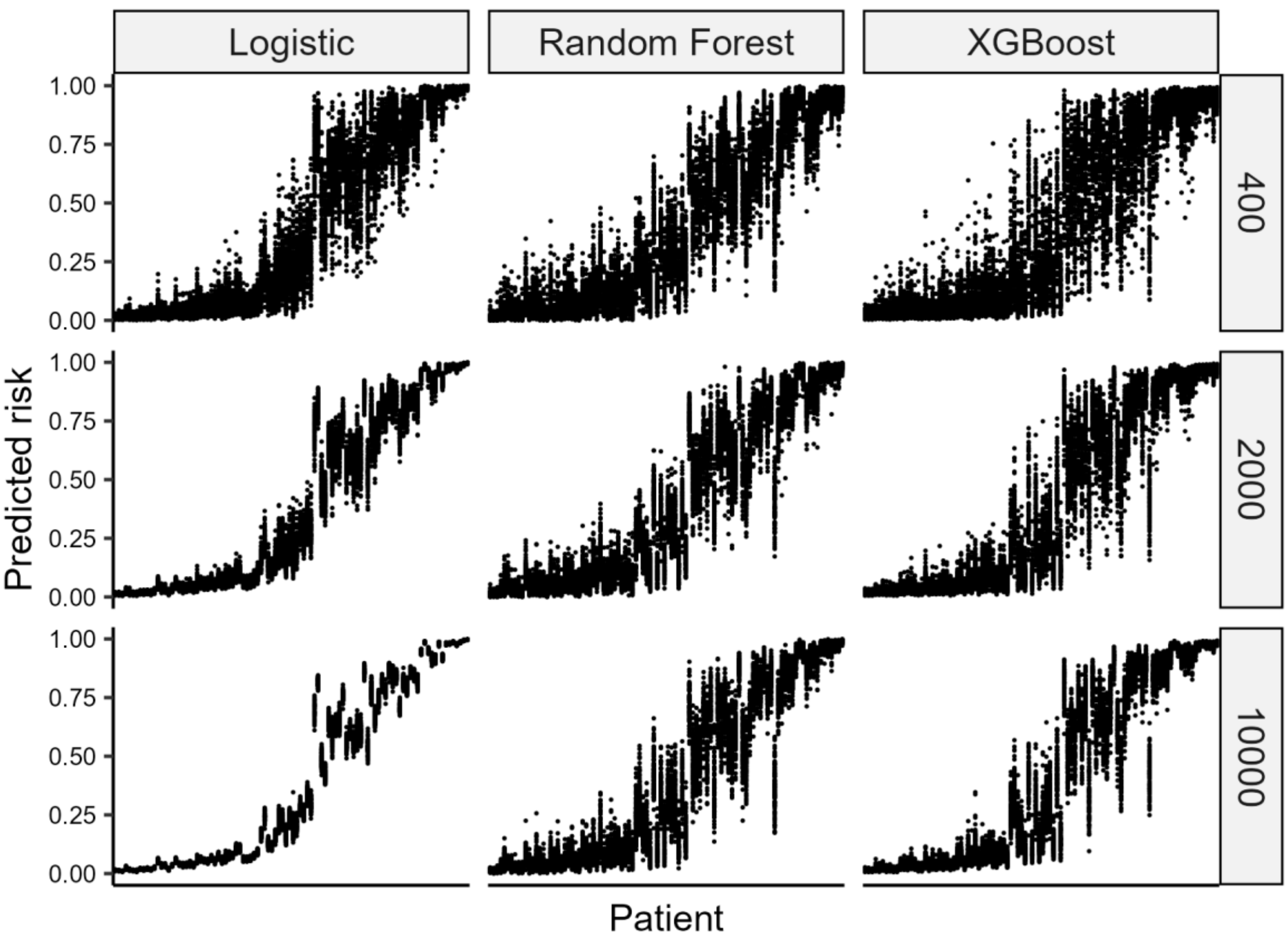

**Fig. S4. Scatterplot of individual risk by patient to illustrate model and modeller uncertainty in synthetic data by type of model and training sample size. Each patient has 33 individual risk estimates from 27 logistic models and 6 tree-based models.**

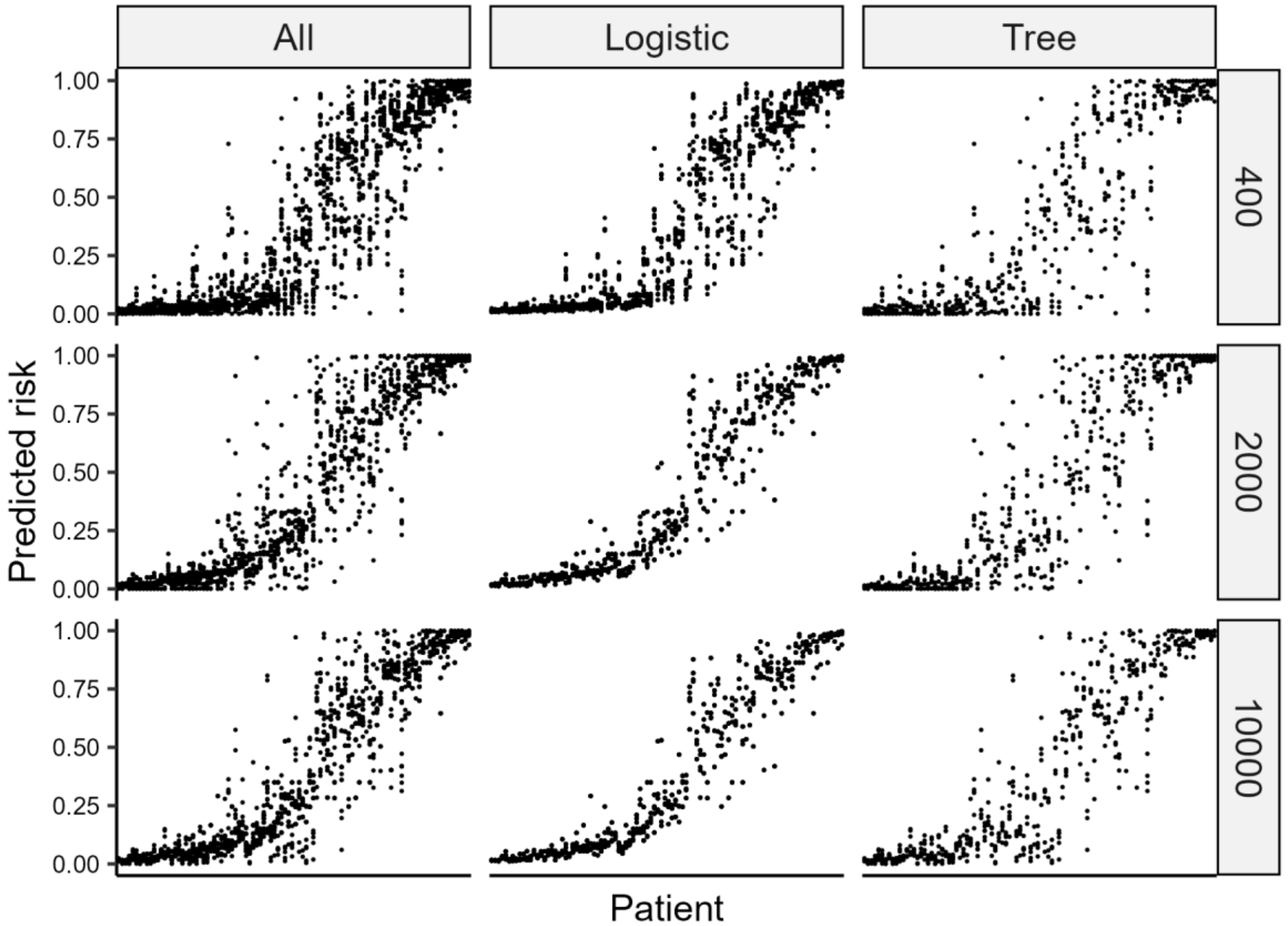

**Fig. S5. Scatterplot of individual risk by patient to illustrate data uncertainty in synthetic data by type of model and training sample size. Each patient has 6 individual risk estimates.**

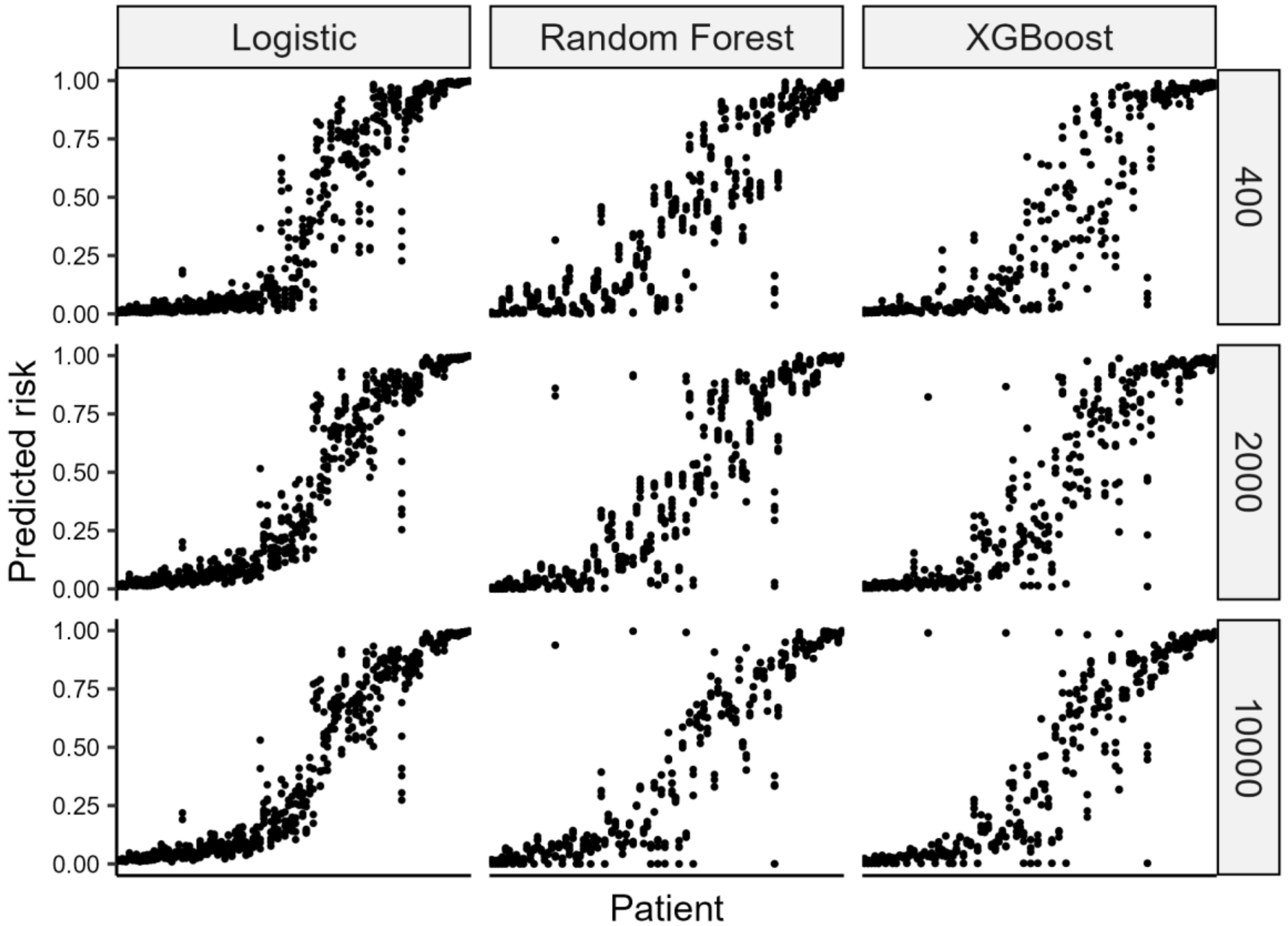

**Fig. S6. Scatterplot of individual risk by patient to illustrate population uncertainty in synthetic data by type of model and training sample size. Each patient has 3 different individual risk estimates depending on the training population.**

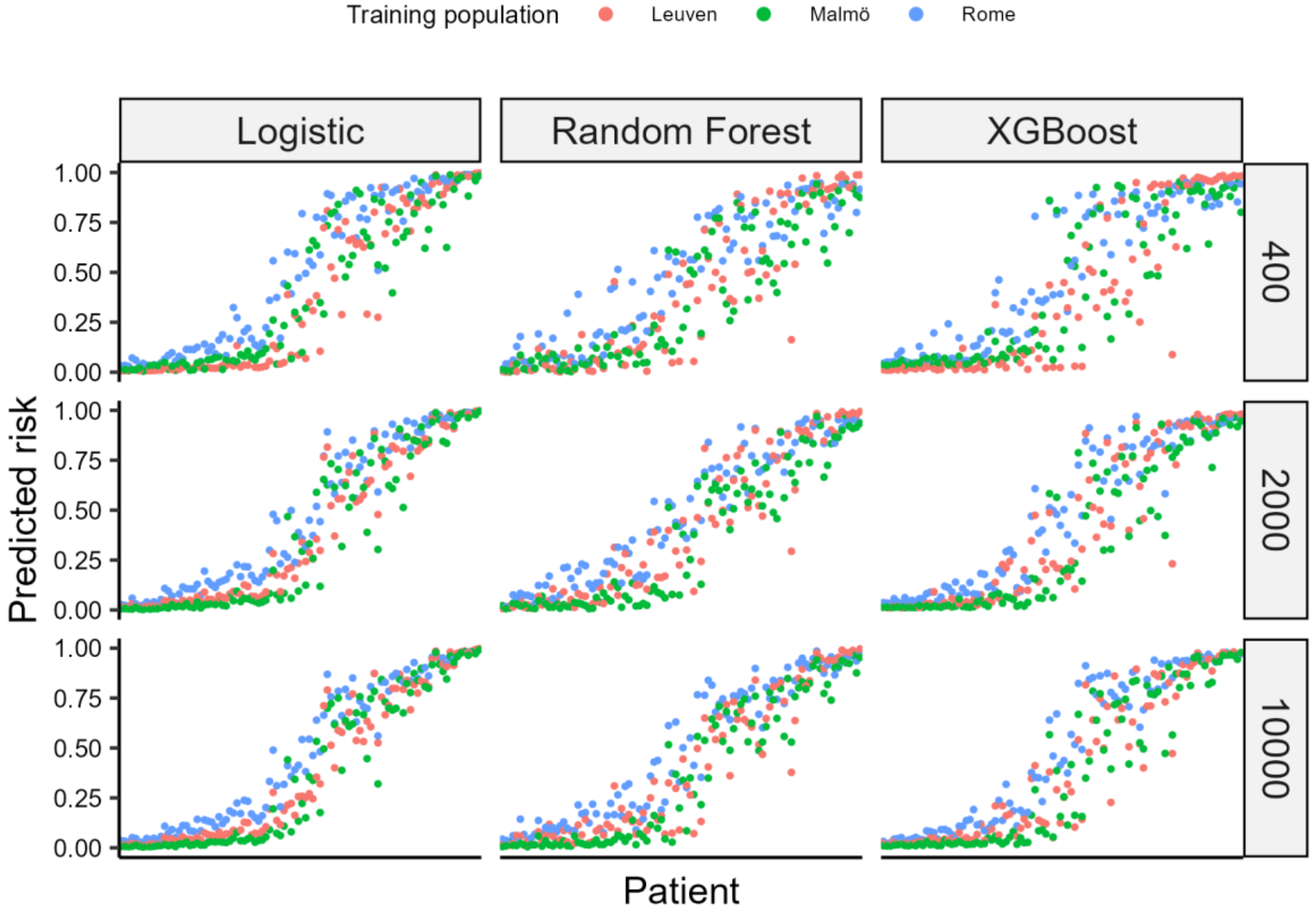

**Fig. S7. Scatterplot of individual risk by patient to illustrate estimation uncertainty in real data (N =400). Each patient has 100 individual risks generated by models where the only variation was training sample.**

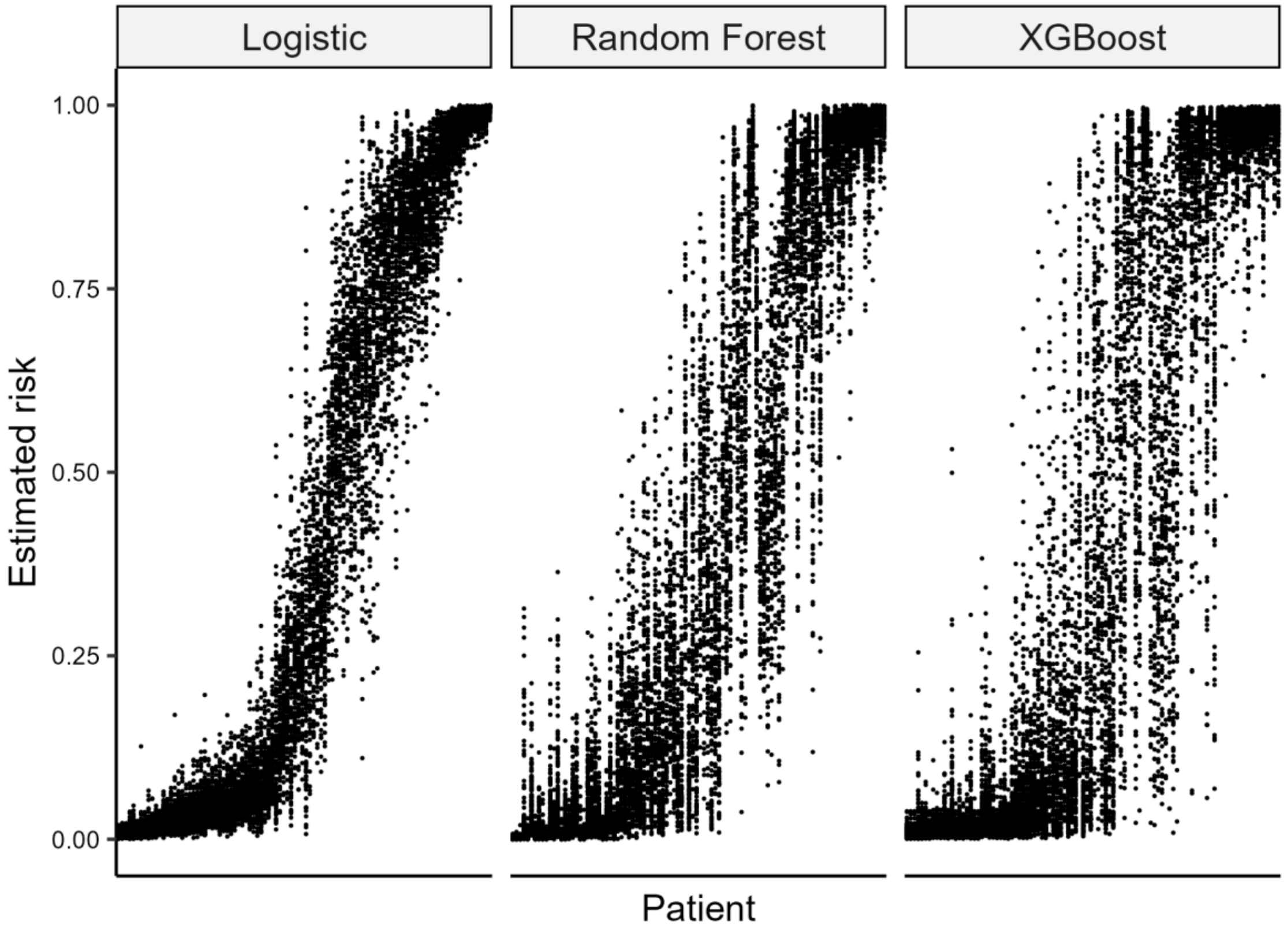

**Fig. S8. Scatterplot of individual risk by patient to illustrate model uncertainty in real data (N =400). Each patient has 33 individual risks in total with 27 in logistic models and 6 in tree based models.**

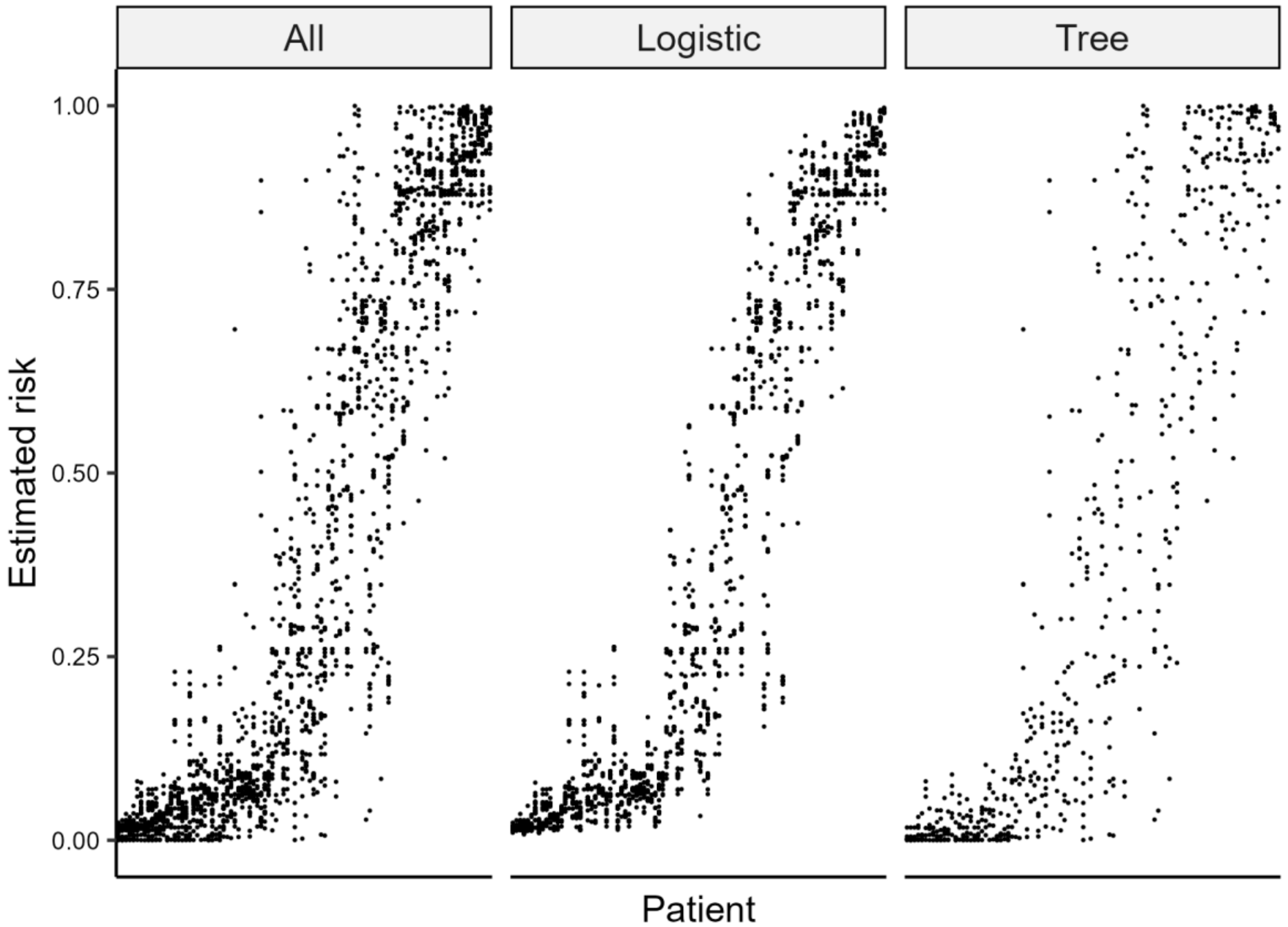

**Fig. S9. Scatterplot of individual risk by patient to illustrate data uncertainty in real data (N =400). Each patient has 6 individual risks.**

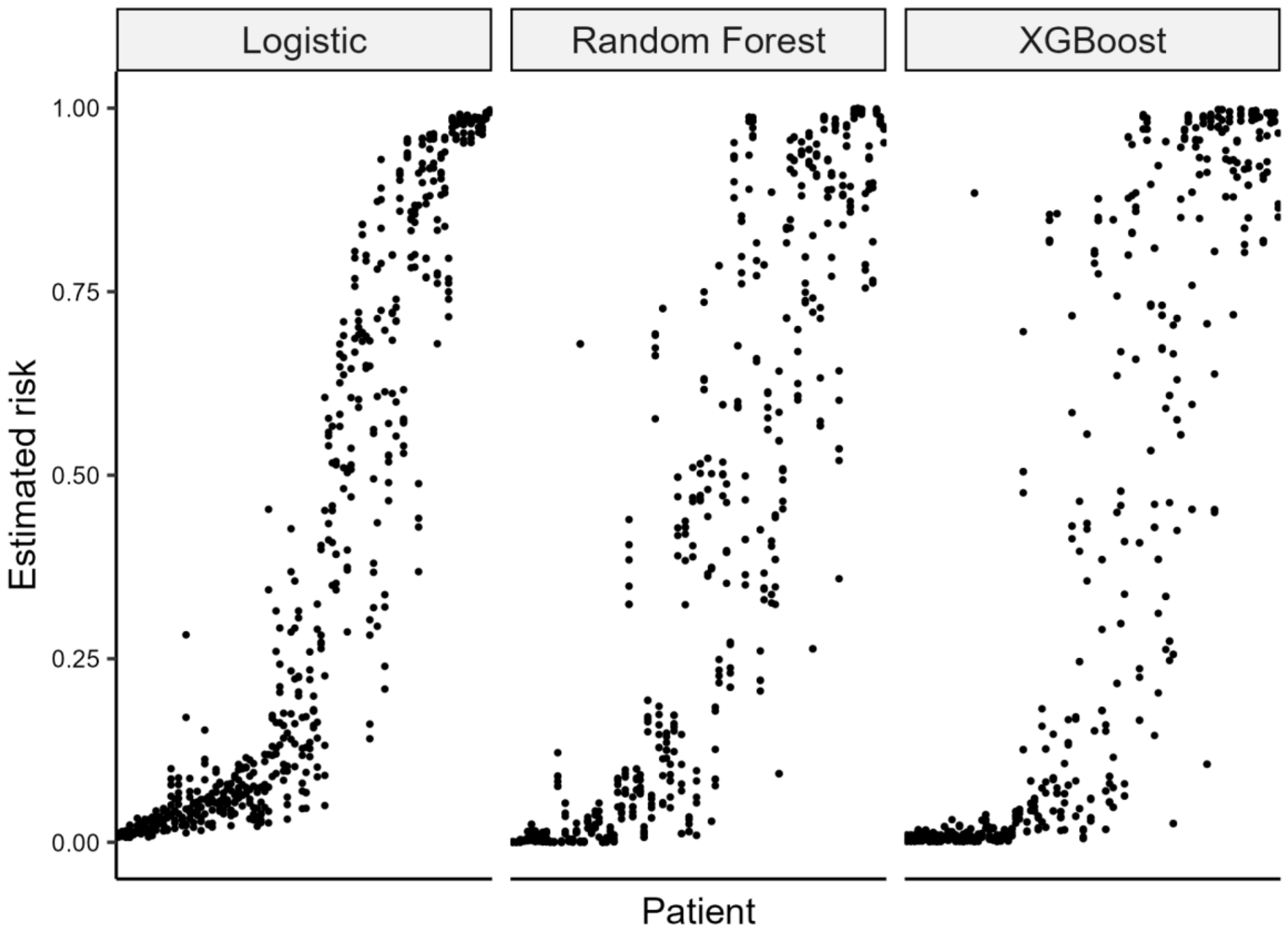

**Fig. S10. Scatterplot of individual risk by patient to illustrate population uncertainty in real data (N =400). Each patient has 3 different individual risks depending on the training population.**

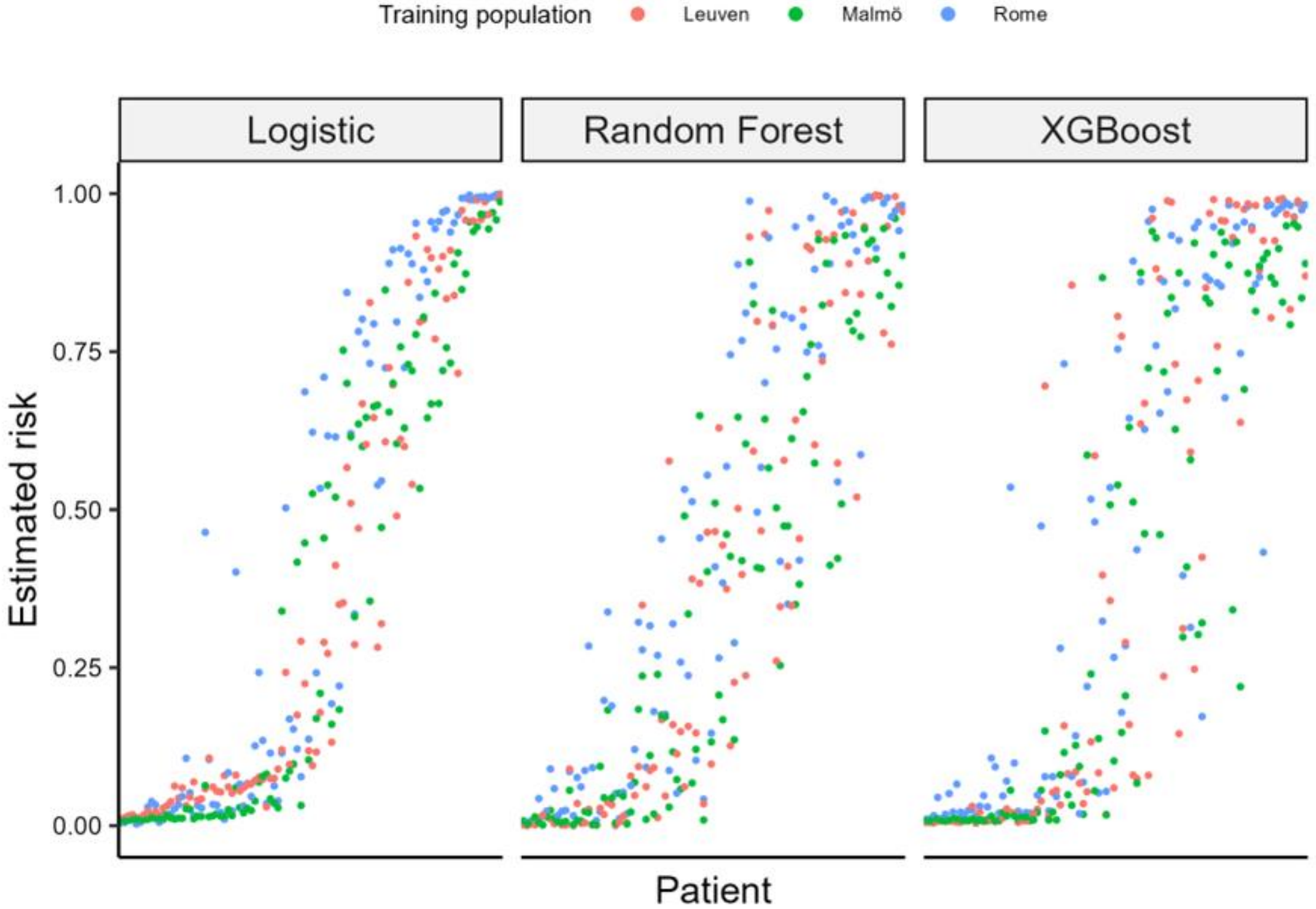

**Table S1. Uncertainty category variations in the simulation study.**

| Category of uncertainty | Variation | Options[a] |
|---|---|---|
| Model | Modeling algorithm | Logistic regression (LR) |
| | | Random forest (RF) |
| | | Extreme gradient boosting (XGB) |
| | Handling of continuous variables (LR only) | Linear |
| | | Dichotomization at median |
| | | Categorization in 4 groups using quartile split |
| | | Multivariable fractional polynomials (MFP) |
| | | Restricted cubic splines with 3 knots |
| | Variable selection (LR only) | None |
| | | Backward elimination with alpha 0.01 |
| | | Backward elimination with alpha 0.20 |
| | Penalization (LR only) | No |
| | | Ridge, lambda tuned using AIC[b] |
| | Minimum node size (RF only)/ Maximum depth (XGB only)[c] | 2 |
| | | 20 |
| | | Tuned (On logloss) |
| Applicability (Data) | Definition of size of lesion and solid component | Based on maximum diameters |
| | | Based on volumes |
| | Handling of missing data for CA125 | Regression imputation |
| | | Median imputation conditional on outcome |
| | | Missing indicator method |
| Applicability (Population) | Training data population | Leuven (Belgium) |
| | | Malmö (Sweden) |
| | | Rome (Italy) |
| Estimation | Training data sample | 100 randomly sampled training datasets (real data only) |
| | Training data size | 400 (real data or synthetic data) |
| | | 2000 (synthetic data only) |
| | | 10000 (synthetic data only) |

[a] The main option is underlined. The main modeling strategy is logistic regression with restricted cubic splines, no variable selection or penalization, using maximum diameters to define size, using regression imputation, and using training data from Leuven. The main RF strategy is fully tuned and uses the same options for applicability uncertainty. The main XGB strategy is analogous: fully tuned and using the same options for applicability uncertainty.

[b] Not used when handling continuous variables using MFP.

[c] For RF, when tuned, we tuned also the hyperparameters sampling fraction and mtry. For XGB, when tuned ,we always tuned the hyperparameters eta, and subsample.

**Table S2. Center specific variable distribution in synthetically generated dataset. For continuous variables, median (interquartile range) and range are shown. For binary variables, n (%) is shown.**

| | Training populations (synthetic) | | | |
|---|---|---|---|---|
| | **Leuven (N = 100000)** | **Malmö (N = 100000)** | **Rome (N = 100000)** | **Leuven Test set (N = 100)** |
| Age (years) | 51 (37-63), 18-88 | 49 (35-63), 18-96 | 50 (39-63), 18-90 | 52 (40-62), 18-84 |
| Volume of lesion (ml) | 97 (29-494), 0.5184-16379.5 | 98 (44-292), 0.4-11347.4 | 150 (38-509), 0.2492-10221 | 166 (37-690), 6.5-11100.8 |
| Maximum diameter of lesion (mm) | 71 (46-113), 11-410 | 71 (52-103), 10-435 | 79 (51-118), 13-320 | 82 (51-127), 26-338 |
| CA125 (IU/L) | 41 (15-270), 3-40140 | 22 (14-59), 2-9035 | 72 (16-311), 2-57900 | 46 (18-224), 6-6203 |
| Missing CA125 | 30222 (30%) | 18334 (18%) | 52960 (53%) | 26 (26%) |
| Prop. of solid tissue based on volume | 0 (0-0.25), 0-1 | 0 (0-0.01), 0-1 | 0.01 (0-1), 0-1 | 0.01 (0-0.38), 0-1 |
| Prop. of solid tissue based on diameter | 0.22 (0-0.61), 0-1 | 0.12 (0-0.4), 0-1 | 0.38 (0-1), 0-1 | 0.21 (0-0.68), 0-1 |
| Bilateral lesions | 18923 (19%) | 23383 (23%) | 24269 (24%) | 30 (30%) |
| Papilary flow | 19680 (20%) | 12123 (12%) | 10833 (11%) | 17 (17%) |
| Malignant tumors | 36727 (37%) | 22566 (23%) | 50292 (50%) | 49 (49%) |

**Table S3. Summary of individual risk uncertainty and model performance in the test set for different uncertainties using the synthetically generated data as training.**

| Uncertainty source | Model type | Training size | 95% range, mean (range) | AUROC, mean (range) | ECI, mean (range) | RU, median (range) | DU |
|---|---|---|---|---|---|---|---|
| Estimation | LR | 400 | 0.22 (0.01; 0.65) | 0.92 (0.89; 0.93) | 0.11 (0.02; 0.21) | 0.05 (-0.61; 0.49) | 0.06 (0; 0.49) |
| | | 2000 | 0.10 (0.01; 0.36) | 0.92 (0.91; 0.93) | 0.11 (0.06; 0.14) | 0.20 (-0.02; 0.49) | 0.02 (0; 0.33) |
| | | 10000 | 0.04 (<.01; 0.14) | 0.93 (0.92; 0.93) | 0.11 (0.10; 0.13) | 0.18 (0.02; 0.37) | 0.01 (0; 0.31) |
| | RF | 400 | 0.26 (0.03; 0.51) | 0.91 (0.89; 0.94) | 0.12 (0.04; 0.26) | 0.08 (-0.35; 0.49) | 0.10 (0; 0.49) |
| | | 2000 | 0.19 (0.02; 0.42) | 0.92 (0.90; 0.94) | 0.10 (0.06; 0.32) | 0.12 (-0.14; 0.49) | 0.07 (0; 0.48) |
| | | 10000 | 0.17 (0.02; 0.41) | 0.93 (0.91; 0.95) | 0.10 (0.06; 0.16) | 0.16 (-0.08; 0.51) | 0.06 (0; 0.49) |
| | XGBoost | 400 | 0.31 (0.04; 0.73) | 0.91 (0.88; 0.94) | 0.11 (0.02; 0.29) | 0.03 (-0.47; 0.51) | 0.10 (0; 0.49) |
| | | 2000 | 0.20 (0.02; 0.51) | 0.93 (0.88; 0.95) | 0.10 (0.04; 0.21) | 0.15 (-0.18; 0.53) | 0.06 (0; 0.47) |
| | | 10000 | 0.14 (0.01; 0.38) | 0.93 (0.92; 0.94) | 0.10 (0.06; 0.20) | 0.18 (-0.12; 0.49) | 0.05 (0; 0.49) |
| Model | All | 400 | 0.22 (0.02; 0.70) | 0.91 (0.81; 0.95) | 0.11 (0.02; 0.39) | 0.14 (-1.96; 0.51) | 0.06 (0; 0.48) |
| | | 2000 | 0.20 (0.01; 0.69) | 0.92 (0.85; 0.95) | 0.10 (0.02; 0.32) | 0.22 (-1.63; 0.53) | 0.05 (0; 0.48) |
| | | 10000 | 0.18 (0.01; 0.67) | 0.92 (0.84; 0.95) | 0.09 (0.03; 0.20) | 0.22 (-1.37; 0.53) | 0.05 (0; 0.48) |
| | Regression | 400 | 0.14 (0.01; 0.52) | 0.91 (0.86; 0.93) | 0.11 (0.02; 0.29) | 0.18 (-0.69; 0.49) | 0.05 (0; 0.48) |
| | | 2000 | 0.11 (0.01; 0.45) | 0.91 (0.88; 0.93) | 0.10 (0.04; 0.17) | 0.22 (-0.22; 0.49) | 0.04 (0; 0.48) |
| | | 10000 | 0.11 (0.01; 0.44) | 0.92 (0.89; 0.93) | 0.09 (0.07; 0.13) | 0.22 (0.02; 0.39) | 0.04 (0; 0.48) |
| | Tree based | 400 | 0.18 (0.02; 0.65) | 0.91 (0.81; 0.95) | 0.12 (0.02; 0.39) | 0.00 (-1.96; 0.51) | 0.07 (0; 0.50) |
| | | 2000 | 0.16 (0.01; 0.64) | 0.92 (0.85; 0.95) | 0.10 (0.02; 0.32) | 0.08 (-1.63; 0.53) | 0.06 (0; 0.50) |
| | | 10000 | 0.13 (0.01; 0.59) | 0.92 (0.84; 0.95) | 0.10 (0.03; 0.20) | 0.14 (-1.37; 0.53) | 0.05 (0; 0.50) |
| Data | LR | 400 | 0.11 (<.01; 0.62) | 0.93 (0.89; 0.96) | 0.08 (0.01; 0.21) | 0.20 (-0.63; 0.65) | 0.05 (0; 0.50) |
| | | 2000 | 0.10 (<.01; 0.60) | 0.94 (0.91; 0.96) | 0.08 (0.02; 0.14) | 0.25 (-0.14; 0.63) | 0.04 (0; 0.50) |
| | | 10000 | 0.09 (<.01; 0.59) | 0.94 (0.92; 0.96) | 0.08 (0.05; 0.13) | 0.25 (0.02; 0.59) | 0.04 (0; 0.50) |
| | RF | 400 | 0.08 (0.01; 0.63) | 0.93 (0.88; 0.97) | 0.13 (0.04; 0.27) | 0.16 (-0.35; 0.82) | 0.04 (0; 0.50) |
| | | 2000 | 0.08 (0.01; 0.86) | 0.94 (0.90; 0.98) | 0.10 (0.04; 0.32) | 0.20 (-0.20; 0.80) | 0.04 (0; 0.50) |
| | | 10000 | 0.08 (<.01; 0.93) | 0.94 (0.91; 0.98) | 0.09 (0.03; 0.17) | 0.26 (-0.18; 0.84) | 0.04 (0; 0.50) |
| | XGBoost | 400 | 0.12 (0.01; 0.73) | 0.93 (0.87; 0.97) | 0.11 (0.02; 0.32) | 0.12 (-0.61; 0.78) | 0.05 (0; 0.50) |
| | | 2000 | 0.11 (0.01; 0.82) | 0.94 (0.88; 0.98) | 0.09 (0.02; 0.23) | 0.24 (-0.25; 0.82) | 0.05 (0; 0.50) |
| | | 10000 | 0.10 (<.01; 0.94) | 0.94 (0.92; 0.98) | 0.09 (0.03; 0.20) | 0.22 (-0.14; 0.82) | 0.04 (0; 0.50) |
| Population | LR | 400 | 0.15 (0.01; 0.54) | 0.92 (0.89; 0.93) | 0.10 (<.01; 0.28) | 0.09 (-0.76; 0.57) | 0.08 (0; 0.33) |
| | | 2000 | 0.12 (<.01; 0.40) | 0.92 (0.91; 0.93) | 0.09 (0.01; 0.18) | 0.20 (-0.25; 0.49) | 0.08 (0; 0.33) |
| | | 10000 | 0.11 (<.01; 0.36) | 0.92 (0.92; 0.93) | 0.09 (0.02; 0.16) | 0.18 (-0.10; 0.45) | 0.08 (0; 0.33) |
| | RF | 400 | 0.17 (0.01; 0.52) | 0.90 (0.85; 0.94) | 0.14 (0.01; 0.64) | 0.08 (-0.59; 0.49) | 0.10 (0; 0.33) |
| | | 2000 | 0.15 (0.01; 0.47) | 0.91 (0.87; 0.94) | 0.10 (0.01; 0.32) | 0.10 (-0.49; 0.49) | 0.10 (0; 0.33) |
| | | 10000 | 0.15 (0.01; 0.49) | 0.92 (0.89; 0.95) | 0.09 (<.01; 0.21) | 0.14 (-0.33; 0.51) | 0.09 (0; 0.33) |
| | XGBoost | 400 | 0.19 (0.01; 0.66) | 0.90 (0.84; 0.94) | 0.12 (<.01; 0.45) | 0.08 (-0.82; 0.53) | 0.10 (0; 0.33) |
| | | 2000 | 0.15 (0.01; 0.55) | 0.92 (0.88; 0.95) | 0.09 (<.01; 0.28) | 0.14 (-0.43; 0.53) | 0.08 (0; 0.33) |
| | | 10000 | 0.14 (0.01; 0.49) | 0.93 (0.90; 0.94) | 0.08 (<.01; 0.20) | 0.18 (-0.24; 0.49) | 0.08 (0; 0.33) |

AUROC: area under the receiver operating characteristic curve; ECI: estimated calibration index; RU: relative utility; DU: decision uncertainty; LR: logistic regression; RF: random forest.

**Table S4. Center specific variable distribution in true dataset. For continuous variables, median (interquartile range) and range are shown. For binary variables, n (%) is shown.**

| | Training populations (true) | | | |
|---|---|---|---|---|
| | **Leuven (N = 1022)** | **Malmö (N = 1048)** | **Rome (N = 1131)** | **Leuven Test set (N = 100)** |
| Age (years) | 51 (36-63), 18-88 | 49 (35-63), 18-96 | 50 (39-63), 18-90 | 52 (40-62), 18-84 |
| Volume of lesion (ml) | 101 (31-442), 0.52-16379 | 106 (40-319), 0.4-11347 | 142 (39-489), 0.25-10995 | 166 (37-690), 6.5-11100 |
| Maximum diameter of lesion (mm) | 71 (46-114), 11-410 | 71 (52-102), 10-435 | 79 (51-118), 13-320 | 82 (51-127), 26-338 |
| CA125 (IU/L) | 44 (16-312), 3-40140 | 23 (14-62), 1-9035 | 74 (17-322), 2-57900 | 46 (18-224), 6-6203 |
| Missing CA125 | 312 (31%) | 187 (18%) | 602 (53%) | 26 (26%) |
| Prop. of solid tissue based on volume | 0.01 (0-0.37), 0-1 | 0 (0-0.07), 0-1 | 0.04 (0-1), 0-1 | 0.01 (0-0.38), 0-1 |
| Prop. of solid tissue based on diameter | 0.21 (0-0.73), 0-1 | 0.09 (0-0.43), 0-1 | 0.38 (0-1), 0-1 | 0.22 (0-0.68), 0-1 |
| Bilateral lessions | 193 (19%) | 247 (24%) | 276 (24%) | 30 (30%) |
| Papilary flow | 188 (18%) | 111 (11%) | 96 (8%) | 17 (17%) |
| Malignant tumors | 378 (37%) | 239 (23%) | 576 (51%) | 49 (49%) |

**Table S5. Summary of individual risk uncertainty and model performance in the test set for different uncertainties based on true  data (N = 400).**

| Uncertainty source | Panel | # models | 95% range, mean (range) | AUROC, mean (range) | ECI, mean (range) | RU, median (range) | DU |
|---|---|---|---|---|---|---|---|
| Estimation | Main modeling strategy | 100 | 0.22 (0.01; 0.71) | 0.92 (0.89; 0.93) | 0.10 (0.03; 0.19) | 0.07 (-0.31; 0.49) | 0.04 (0; 0.50) |
| | RF main strategy | 100 | 0.29 (<.01; 0.70) | 0.91 (0.87; 0.93) | 0.08 (0.02; 0.20) | 0.02 (-0.39; 0.45) | 0.1 (0; 0.47) |
| | XGBoost main strategy | 100 | 0.35 (0.02; 0.87) | 0.91 (0.88; 0.94) | 0.10 (0.01; 0.32) | -0.08 (-0.75; 0.51) | 0.09 (0; 0.50) |
| Model | All variations | 33 | 0.22 (0.01; 0.74) | 0.91 (0.84; 0.94) | 0.09 (0.01; 0.32) | 0.08 (-1.90; 0.57) | 0.05 (0; 0.48) |
| | LR only | 27 | 0.14 (0.01; 0.53) | 0.91 (0.85; 0.93) | 0.09 (0.01; 0.24) | 0.14 (-0.69; 0.57) | 0.04 (0; 0.48) |
| | Tree-based only | 6 | 0.18 (0.02; 0.68) | 0.91 (0.84; 0.94) | 0.10 (0.01; 0.32) | 0.00 (-1.90; 0.51) | 0.06 (0; 0.50) |
| Data | LR | 6 | 0.10 (<.01; 0.64) | 0.93 (0.89; 0.96) | 0.08 (0.01; 0.19) | 0.20 (-0.41; 0.67) | 0.04 (0; 0.50) |
| | RF | 6 | 0.08 (<.01; 0.63) | 0.92 (0.87; 0.97) | 0.08 (0.01; 0.21) | 0.22 (-0.41; 0.69) | 0.03 (0; 0.50) |
| | XGBoost | 6 | 0.13 (0.01; 0.80) | 0.92 (0.86; 0.97) | 0.09 (0.01; 0.33) | 0.04 (-0.76; 0.84) | 0.05 (0; 0.50) |
| Population | LR | 3 | 0.15 (<.01; 0.59) | 0.91 (0.88; 0.93) | 0.10 (<.01; 0.40) | 0.10 (-0.59; 0.55) | 0.07 (0; 0.33) |
| | RF | 3 | 0.19 (0.01; 0.66) | 0.90 (0.84; 0.94) | 0.10 (0.01; 0.31) | 0.08 (-0.84; 0.55) | 0.09 (0; 0.33) |
| | XGBoost | 3 | 0.20 (0.01; 0.77) | 0.90 (0.84; 0.94) | 0.10 (0.01; 0.43) | -0.04 (-1.35; 0.57) | 0.08 (0; 0.33) |
| All | All | 59400 | 0.46 (0.05; 0.93) | 0.92 (0.76; 0.97) | 0.11 (0.00; 0.81) | 0.16 (-1.90; 0.84) | 0.11 (0; 0.50) |

AUROC: area under the receiver operating characteristic curve; ECI: estimated calibration index; RU: relative utility; DU: decision uncertainty; LR: logistic regression; RF: random forest.